\documentclass[journal=jacsat,manuscript=article]{achemso}
\usepackage{listings}
\usepackage{lmodern}
\usepackage{amsmath}
\usepackage{amssymb}
\usepackage{todonotes}
\usepackage{graphicx}

\usepackage{xcolor}
\usepackage{soul}
\setkeys{acs}{maxauthors = 0}

\author{Alison~C.~Twitchett-Harrison}
\email{act27@cam.ac.uk}
\author{James~C.~Loudon}
\affiliation{Department of Materials Science and Metallurgy, University of Cambridge, 27 Charles Babbage Road, Cambridge, CB3 0FS, United Kingdom.}
\author{Ryan~A.~Pepper}
\affiliation{Faculty of Engineering and Physical Sciences, University of Southampton, Southampton, SO17 1BJ, United Kingdom.}
\author{Max~T.~Birch}
\affiliation{Max Planck Institute for Intelligent Systems, 70569 Stuttgart, Germany.}
\alsoaffiliation{Department of Physics, Durham University, Durham, DH1 3LE, United Kingdom.}
\author{Hans~Fangohr}
\affiliation{Faculty of Engineering and Physical Sciences, University of Southampton, Southampton, SO17 1BJ, United Kingdom.}
\alsoaffiliation{Max Planck Institute for Structure and Dynamics of Matter, Luruper Chaussee 149, 22761 Hamburg, Germany.}
\author{Paul~A.~Midgley} 
\affiliation{Department of Materials Science and Metallurgy, University of Cambridge, 27 Charles Babbage Road, Cambridge, CB3 0FS, United Kingdom.}
\author{Geetha~Balakrishnan}
\affiliation{Department of Physics, University of Warwick, Coventry, CV4 7AL, United Kingdom.}
\author{Peter~D.~Hatton}
\affiliation{Department of Physics, Durham University, Durham, DH1 3LE, United Kingdom.}
\title{Confinement of Skyrmions in Nanoscale FeGe Device-like Structures}

\begin{document}

\begin{abstract}
Skyrmion-based devices have been proposed as a promising solution for low energy data storage. These devices include racetrack or logic structures and require skyrmions to be confined in regions with dimensions comparable to the size of a single skyrmion. Here we examine skyrmions in {FeGe} device shapes using Lorentz transmission electron microscopy (LTEM) to reveal the consequences of skyrmion confinement in a device-like structure. Dumbbell-shaped elements were created by focused ion beam (FIB) milling to provide regions where single skyrmions are confined adjacent to areas containing a skyrmion lattice.  Simple block shapes of equivalent dimensions were also prepared to allow a direct comparison with skyrmion formation in a less complex, yet still confined, device geometry. The impact of applying a magnetic field and varying the temperature on the formation of skyrmions within the shapes was examined. This revealed that it is not just confinement within a small device structure that controls the position and number of skyrmions, but that a complex device geometry changes the skyrmion behaviour, including allowing skyrmions to form at lower applied magnetic fields than in simple shapes.  The impact of edges in complex shapes is observed to be significant in changing the behaviour of the magnetic textures formed.  This could allow methods to be developed to control both the position and number of skyrmions within device structures.

\end{abstract}

\section{Keywords}
{FeGe}, LTEM, Bloch skyrmions, confinement, FIB-microfabrication, dumbbell-shape, micromagnetic simulations.


\section{Introduction}

Magnetic skyrmions are localised magnetic configurations with an integer, non-zero topological charge. They resemble magnetic vortices and typically have diameters of a few tens of nanometres. Magnetic skyrmions were discovered experimentally in 2009~\cite{Muh09} and since then there have been many suggestions for how they could be used in spintronic devices (reviewed in refs.~\cite{Fert13, Felser13, Kang16, Wiesendanger16, Sitte18, Lancaster19}). Proposed devices include racetrack memories~\cite{Parkin08, Sampaio13, Parkin15, Gobel19, Kang16b}, logic devices~\cite{Zhou14, Zhang15}, skyrmion transitors~\cite{Zhang15a}, nanometre-sized spin transfer oscillators~\cite{Zhang15b,Garcia16} as well as devices for probabilistic~\cite{Pinna18, Zazvorka19}, reservoir~\cite{Bourianoff18} and neuromorphic~\cite{Huang17, Li17a, Grollier16} computing. All of these devices (apart from the reservoir computer) require the skyrmions to be confined in a narrow track with a width a few times the skyrmion diameter. In the case of the logic devices and some schemes for the generation and annihilation of skyrmions~\cite{Iwasaki13, Zhang14}, the skyrmions must be driven through constrictions narrower than the width of a single skyrmion.  Some of these devices use interfacial Dzyalohsinskii-Moriya (DM) interactions to form skyrmions in thin films, others use bulk DM interactions in confined geometries.  The operation of these devices has been simulated but, with the exception of the skyrmion reshuffler~\cite{Zazvorka19}, has not been investigated experimentally. 

In order to create a reliable, reproducible skyrmion device, we must first understand the factors which control the position, number, and size of skyrmions within a material. Devices such as racetrack memories \cite{Fert13} and discrete geometrical shapes have been studied including discs \cite{Zhao16}, wedges \cite{Jin17}, triangles and rectangles \cite{Pepper2018, Matsumoto18}.  These geometries confine the skyrmions and reveal the impact of physical confinement of skyrmions which is key to the development of potential device shapes. Current driven motion of skyrmions in FeGe has also been studied to reveal the behaviour of skyrmions in thin lamellae \cite{Tang21} \cite{Yu20}. However, to date, the control of individual skyrmions within a device structure is one factor that has limited the development of reliable skyrmionic data storage and retrieval \cite{Kang16}. In this paper we use Lorentz transmission electron microscopy (LTEM) to examine  dumbbell-shaped {FeGe} elements, all cut from a single crystal, which confine skyrmions to small regions, including a central constriction with a width comparable to the skyrmion diameter ($70$~nm in {FeGe}).  This forms the type of junction that would need to be used in a skyrmion logic device \cite{Zhang15} or for skyrmion creation and annihilation \cite{Zhang14}.

\section{Results and Discussion}

\subsection{Geometrical confinement of skyrmions in complex shapes}

\begin{figure}[ht!]
    \centering
     \includegraphics[width=13.5cm]{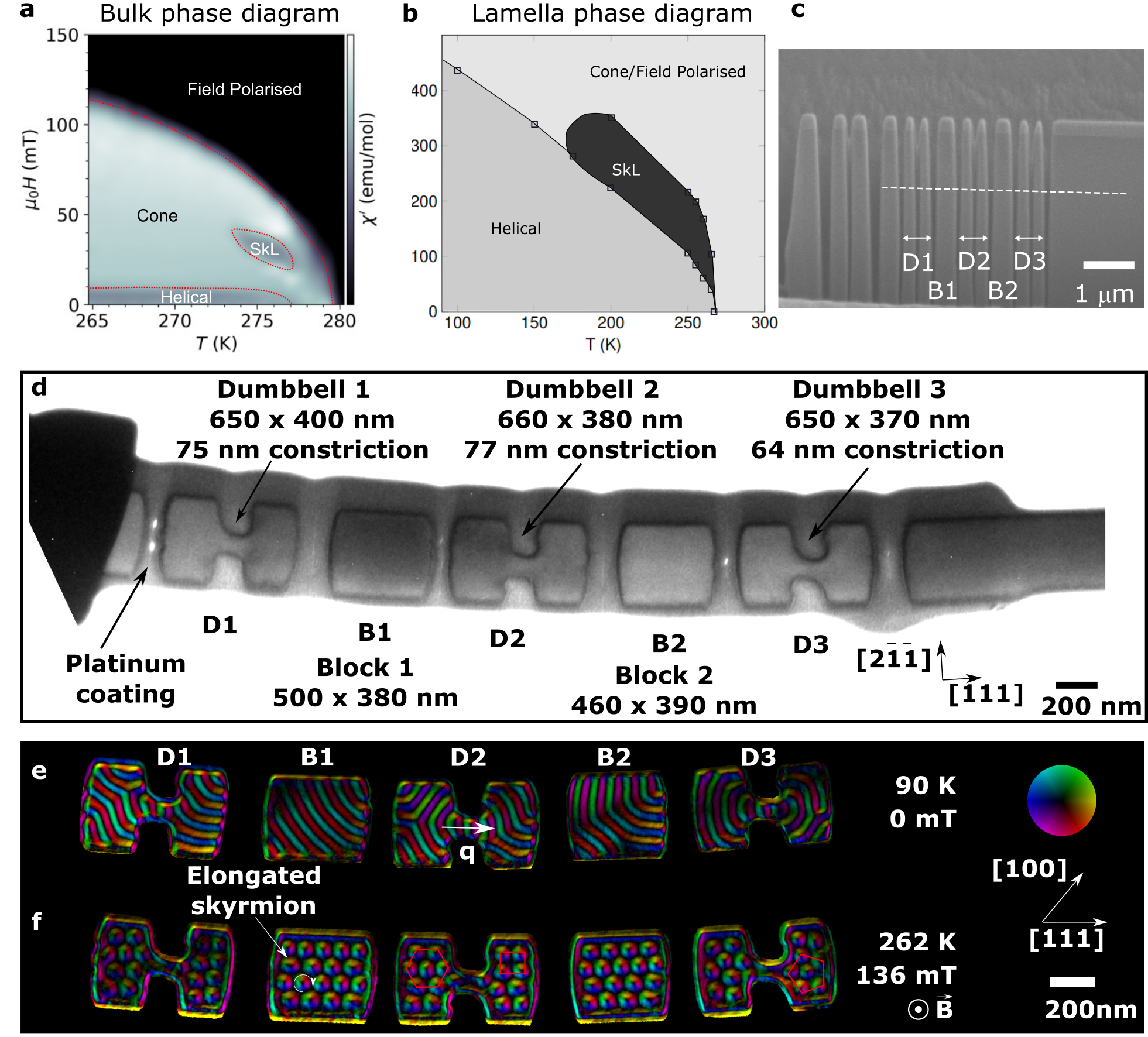}
    \caption{(a) {FeGe} bulk phase diagram from AC susceptibility measurements showing regions of skyrmion lattice (SkL). (b) Phase diagram from small angle x-ray scattering measurements for a $120$~nm thick {FeGe} lamella. These phase diagrams have been adapted from previously published data \cite{Birch20}. (c) Scanning electron micrograph after selective milling of the device-like shapes D1-3 and B1-2 using the focused ion beam. The dashed white line indicates the position and length of the final TEM membrane. (d) Bright field TEM image of the thinned shapes. (e) TIE reconstruction of the magnetic induction of the helical phase in the shapes at $90$~K. The colour wheel indicates the local direction of the in-plane magnetic flux density. (f) Colour map of the TIE reconstructed in-plane magnetic induction of the skyrmion phase from LTEM measurements at 262~K and an applied external magnetic field of $136$~mT. The packing of the skyrmions is dependent on the geometry and ranges from square packing (as seen in D2) to pentagonal (D3) to the usual hexagonal (D2) coordination. Masks have been applied around each shape to show only the in plane magnetic induction in the FeGe shapes from the TIE reconstruction and therefore remove the strong intensity modulations arising from the {FeGe}/{Pt} interface. Data shown in (e) and (f) were acquired directly after cooling the specimen from room temperature (i.e. above the Curie Temperature, T$_C$) in zero field conditions.  Temperature and field conditions were chosen to reflect equilibrium conditions for (e) helical and (f) skyrmion phase formation.} 
    \label{fig:Fig1}
\end{figure}

Creating a thin lamella from bulk material is known to increase the range of temperature and magnetic field over which skyrmions can occur~\cite{Yu11, Beg15}. Figure~\ref{fig:Fig1}(a) shows the phase diagram of the bulk {FeGe} crystal used in this experiment (adapted from previously published data \cite{Birch20}). The real component of the AC susceptibility, $\chi$', was measured as a function of increasing field at a range of temperatures. Dotted lines show the phase boundaries between the helical, conical, skyrmion lattice (SkL) and field polarised states. Figure~\ref{fig:Fig1}(b) is a phase diagram for a $120$~nm thick lamella of FeGe from the same crystal derived from magnetic x-ray diffraction measurements(adapted from previously published data\cite{Birch20}). It can be observed that the skyrmion `pocket' has been greatly enlarged. This increased range of stability is well known and is related to the modification of the magnetic structure imposed by the specimen surfaces known as `surface twists'~\cite{LeonovPRL16}.  To investigate the effect of further confining lamellae into nano-scale, device-like geometries, focused ion beam (FIB) milling was used to cut discrete elements from a single crystal of FeGe as shown in Figure \ref{fig:Fig1}(c) (see Figure S1 for details). A thin membrane containing these elements was prepared in cross-section and comprises three dumbbell-shaped elements (D1, D2 and D3). These dumbbells are separated by approximately rectangular blocks (B1 and B2) to allow comparison of the two shapes and examine the impact of the dumbbell-shape on skyrmion formation. The sides of the elements were coated with platinum which was electron beam deposited to connect and protect them during sample preparation (as shown in Figure \ref{fig:Fig1}(d)), and schematically in Figure S1). In the finished TEM specimen, a platinum layer is only present to the sides of the elements and not above or below the {FeGe} device-like structures in the electron beam direction. 

Whilst the blocks and dumbbells were cut from single crystal {FeGe}, the weaker Pt layer surrounding the isolated shapes creates a subtle orientation variation between the shapes, and the resulting diffraction contrast cannot be minimised in all of the shapes. This diffraction contrast is visible in the bright field LTEM image in Figure~\ref{fig:Fig1}(d) where some of the shapes appear brighter (B2 and D3) and some appear darker (B1 and D2) despite the uniform illumination.  There are also some areas of contrast within each isolated shape (for example, on the right hand side of D1) that are likely to arise from local strain caused by sample preparation.  Strain is known to modify the DMI locally \cite{Shibata15} and can stabilise the skyrmion phase \cite{Butenko10}.  The impact of strain in these small structures may enhance skyrmion formation, but the dominating factor on skyrmion shape and formation is expected to be geometrical confinement \cite{Jin17}. Further information on the image processing used to reduce the impact of diffraction contrast on the observed magnetic contrast can be found in the Materials and Methods section. Ultimately, a device of this, or a similar design, may allow the movement of skyrmions to be controlled by `squeezing' the skyrmion through the constriction using an applied electrical current or magnetic field gradient \cite{Iwasaki13, Zhang18}, with the prospect of allowing control of the speed and number of skyrmions within different areas of a device structure.

 Figure~\ref{fig:Fig1}(e) shows the entire magnetic structure imaged at $90$~K in field-free conditions. The colors indicate the in-plane component of the magnetic flux density reconstructed from electron micrographs using the transport of intensity equation (TIE)\cite{Beleggia04}. The helical phase filled all of the device-like shapes but near the central constrictions, the wave-vector of the helix ${\bf q}$ rotated to run parallel to the edges of the constrictions (as indicated by an arrow in D2). At $262$~K in an applied magnetic field of $136$~mT, skyrmions formed in all of the device-like shapes as expected from the phase diagram shown in Figure \ref{fig:Fig1}(b). Figure \ref{fig:Fig1}(f) shows a colour map of the TIE-reconstructed in-plane magnetic flux density. In the block shapes (B1 and B2), the skyrmions form an approximately hexagonal lattice aligned with the longer axis of the shape, but with elongated skyrmions where confinement restricts the number of skyrmions that can fit in the device-like shape (as labelled in B1). We can see in Figure \ref{fig:Fig1}(f) that the dumbbells are not filled with a regular close-packed hexagonal arrangement of skyrmions as observed in the blocks and in larger FeGe samples \cite{Yu11, Kovacs16}. In dumbbells D1 and D2, the skyrmions in the smaller ends to the right of the constrictions form an approximately square-lattice arrangement whereas in the larger dumbbell ends the arrangement is closer to hexagonal, as shown in red. In D3, the smallest of the dumbbells, the arrangement of skyrmions is best described as `space-filling' because confinement does not allow the formation of more than two skyrmions along the top and bottom edges of the dumbbell ends, and arrangements with an irregular $5$-fold coordination can be seen, one of which is outlined in red. These observations indicate that skyrmions in the confined shapes near the Curie temperature (T$_C$) use irregular particle-like packing arrangements, filling and adapting to the space available rather than forming regular hexagonal lattices. This includes within the narrow constrictions in the dumbbell shapes D1 and D2 where skyrmions form at the edges of the constriction but are distorted. Skyrmions are not observed in the narrow constriction in D3 which, at 64~nm, appears to be too narrow to allow the formation of skyrmions.  

\subsection{Hysteretic behaviour of skyrmion formation in {FeGe} shapes}

\begin{figure}[ht]
    \centering
    \includegraphics[width = 1\textwidth]{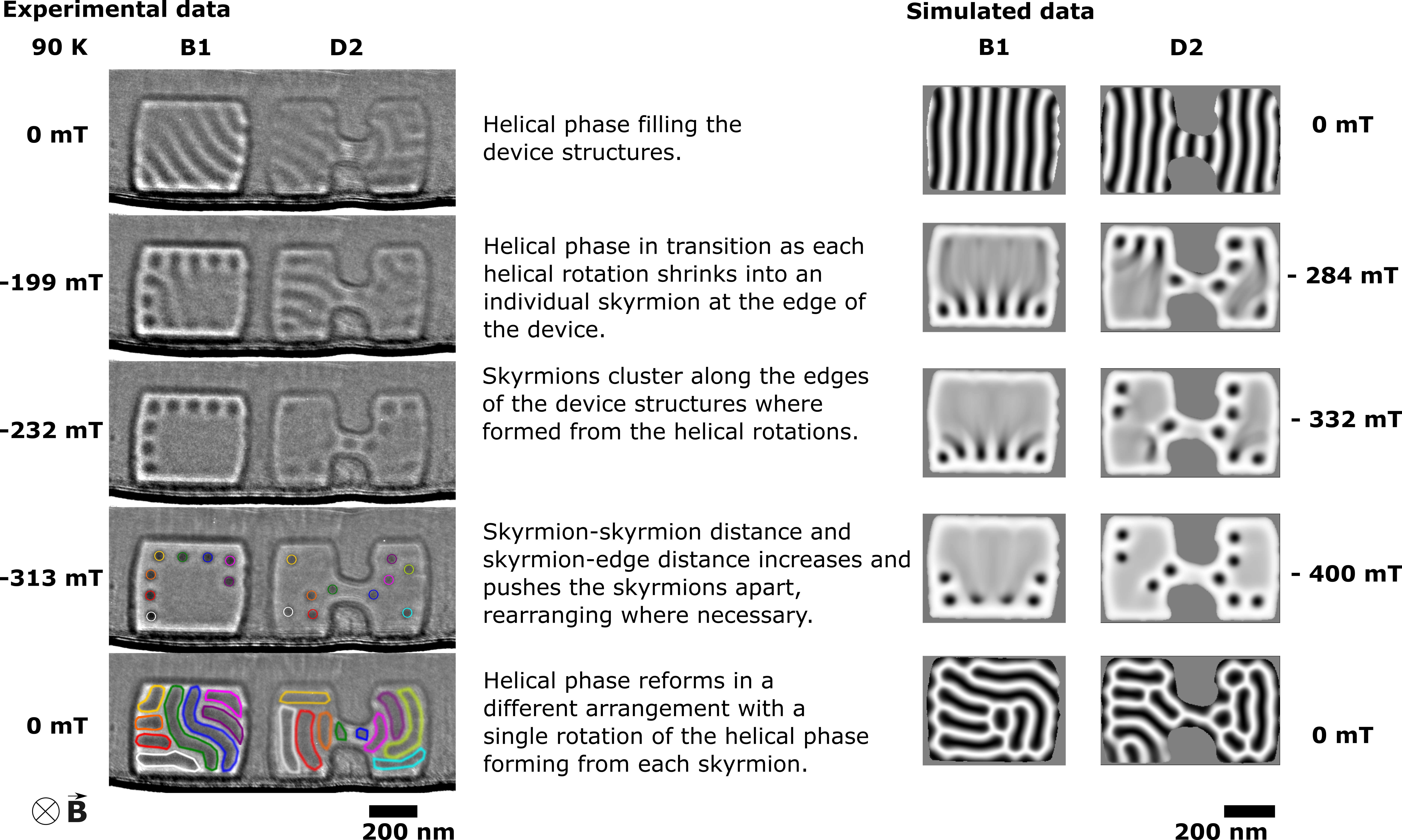}
    \caption{Selected LTEM images acquired at $90$~K.  A direct correlation can be made between skyrmions present at high applied magnetic field ($-313$~mT, marked with coloured circles) and the helical rotations present at $0$~mT (with corresponding coloured lines). Micromagnetic simulations showing the magnetisation of the device-like  shapes reveal a strong similarity to the experimental data.}
    \label{fig:90K}
\end{figure}

\begin{figure}[ht!]
    \centering
    \includegraphics[width = 0.75\textwidth]{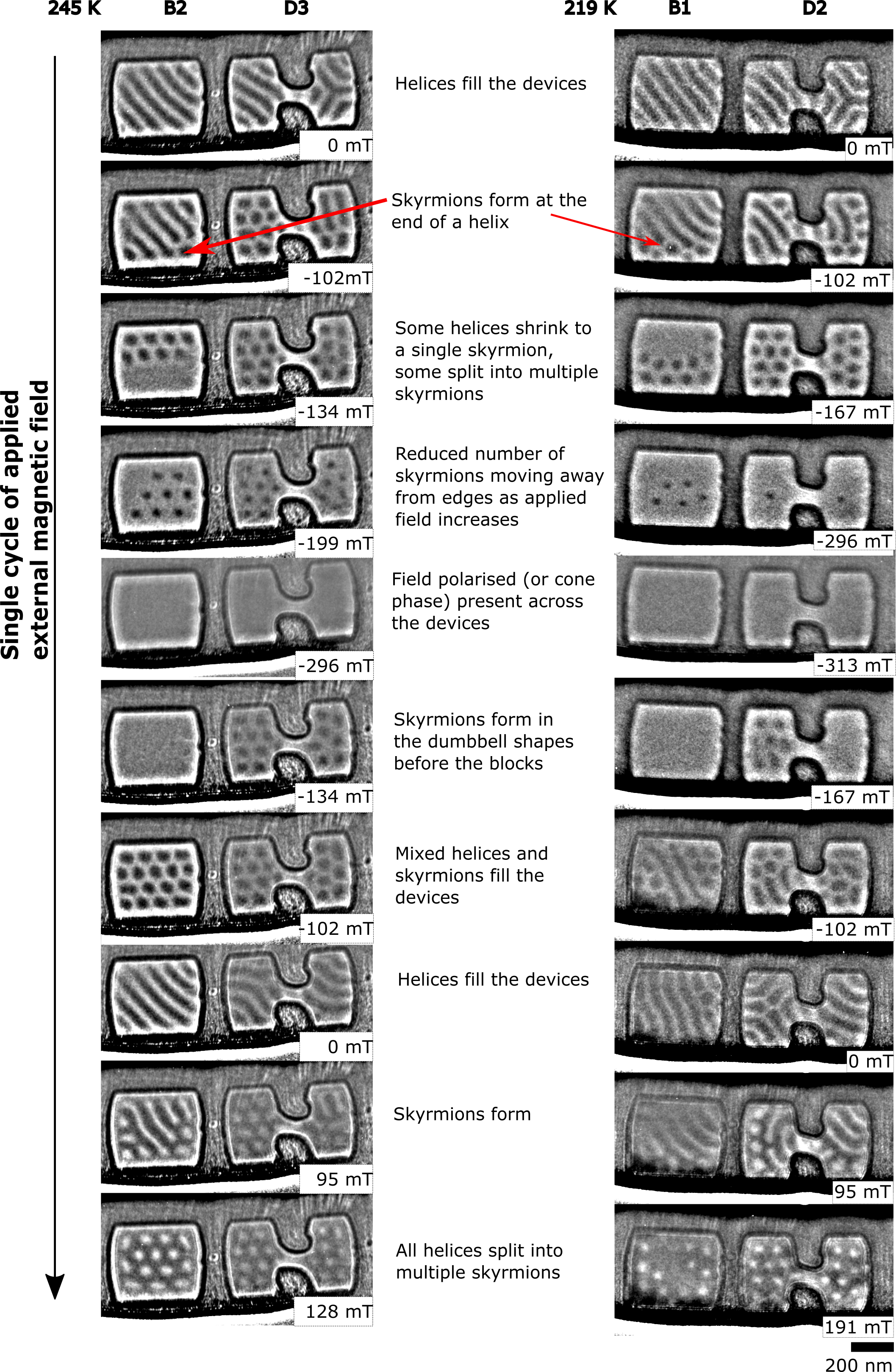}
    \caption{Selected images from the hysteresis loop series acquired at 245~K and 219~K revealing a dependence of observed skyrmion formation on the geometry of the confined shape. The strong diffraction contrast under some values of applied magnetic field (for example, $-102$~mT) reduces the observed magnetic contrast in D3 when compared to the contrast seen in B2.}
    \label{fig:245_219K}
\end{figure}

To investigate the impact of a magnetic field on the magnetic microstructure formed within the device-like shapes, LTEM images were acquired in an applied magnetic field that was varied from $0$~mT$\rightarrow$~$-313$mT~$\rightarrow$~$+310$~mT~$\rightarrow$~$0$~mT. These hysteresis experiments were conducted at $90$~K, $219$~K and $245$~K. Figures \ref{fig:90K} and \ref{fig:245_219K} show a selection of enhanced experimental images (see Materials and Methods) from the hysteresis loops (see Figures S3-8 for full data sets). The phase diagram in Figure \ref{fig:Fig1}(b) shows that any skyrmions formed are metastable at 90 K in thinned TEM lamellae \cite{Yu10}. In these experimental images we observe that skyrmions form from helices in two different ways depending on the temperature. At $90$~K (Figure \ref{fig:90K}), each helical rotation shrinks along its length to form a single skyrmion whereas at $219$~K and $245$~K (Figure \ref{fig:245_219K}) each helical rotation breaks into multiple skyrmions. At $90$~K, the skyrmions formed sit close to the edges of the FeGe shapes, including in the constrictions of the dumbbell shapes, and they rearrange themselves along the edges as the applied magnetic field is increased, spacing out slightly under a $313$~mT applied magnetic field. On reduction of the applied magnetic field, the process is reversed but the skyrmions do not move back into their original positions. A single helical rotation nucleates from each skyrmion at its new position, thereby creating a modified helical arrangement in the small device-like shapes. The skyrmions gather along the edges in lines within the dumbbell shapes in Figure \ref{fig:90K}. This configuration is likely to be formed as a minimisation of energy under the conditions where skyrmions have an attractive force between them \cite{Du15}, and at low temperature where the skyrmion-edge distance is minimised\cite{LeonovPRL16}.  The simulated data shows a strong similarity to the experimental data, following the observed patterns of helices shrinking into single skyrmions, rearranging along the edges of the device-like shapes with increasing field and then the re-forming of helical rotations in more complex configurations as a result of the rearrangement of the skyrmions at high field.  This is observed for both the block and dumbbell shapes. The micromagnetic simulations show the greatest qualitative similarity to the experimental results at low temperatures, where skyrmions are not expelled entirely from the samples at the fields reached in this study (up to ~$310$~mT), while in the micromagnetic simulations skyrmions are present up to much higher applied fields, not being entirely expelled even at $800$~mT (see Zenodo data repository\cite{Zenodo} for full data set). We believe that this behaviour can be explained by a combination of the effects of thermal variations, edge roughness of the real samples, and magnetocrystalline anisotropy which are not explicitly considered in the simulations. 

These results provide an important insight into skyrmion behaviour which is critical for the design of skyrmion devices. The temperature and field history, in combination with device geometry, could be used to predict skyrmion positions and the inter-relationship between helical phase and skyrmion number at low temperature would allow control of skyrmion density within a given device geometry. Figure \ref{fig:245_219K} shows data from the hysteresis loops at $219$~K and $245$~K with the selected shapes chosen as those least affected by diffraction contrast throughout the hysteresis loop, passing through the region of the phase diagram (Figure 1(b)) for thermodynamically stable skyrmion formation. Here we see that, when an external magnetic field is applied,  a skyrmion lattice is formed in the device-like shapes, which partially or completely fills the FeGe shapes. Partial filling of a material with skyrmions is suggested to be due to the energy barrier in a real material\cite{Jin17}\cite{Cortes17}.  It is seen in larger standard FIB-prepared TEM membranes \cite{Li17} and therefore is not expected to be as a result of confinement. The skyrmions formed fill (or partially fill) the shapes and as the applied external magnetic field is increased, slowly reduce in number and move away from the edges. The correlation between helical phase and skyrmion formation is not as well defined at these higher temperatures than at $90$~K and thermodynamic contributions likely encourage the splitting of helical rotations into two or more skyrmions, rather than the pattern of shrinking helices observed at $90$~K. These experimental data also reveal a dependence of skyrmion formation on device geometry. Within Figure \ref{fig:245_219K} we can observe the first formation of skyrmions from the field-polarised state in the device-like  shapes as the applied magnetic field is reduced to $167$~mT at $219$~K and to $134$~mT at $245$~K. Here we can see that skyrmions consistently form first in the dumbbell shapes when transitioning from the field-polarised state, although not always filling the dumbbell shapes entirely. The dumbbell shapes have $1.3$ times the edge length to area ratio compared to the block shapes, which reduces the nucleation energy required for skyrmion formation \cite{Muller16}, thereby allowing them to form in the complex dumbbell shapes before the simple blocks (see supporting information for data). To further quantify this effect, analysis using smaller field steps and varying the rate of temperature change is required.  These experimental results indicate that increasing the edge length to area ratio in a chosen device geometry could be used to stabilise skyrmions and reveals that the skyrmion density within a device is dependent on the geometry of the device and not just the temperature and applied magnetic field.

\subsection{Magnetic structure observed under zero field conditions}

\begin{figure}[ht]
    \centering
    \includegraphics[width=0.9\textwidth]{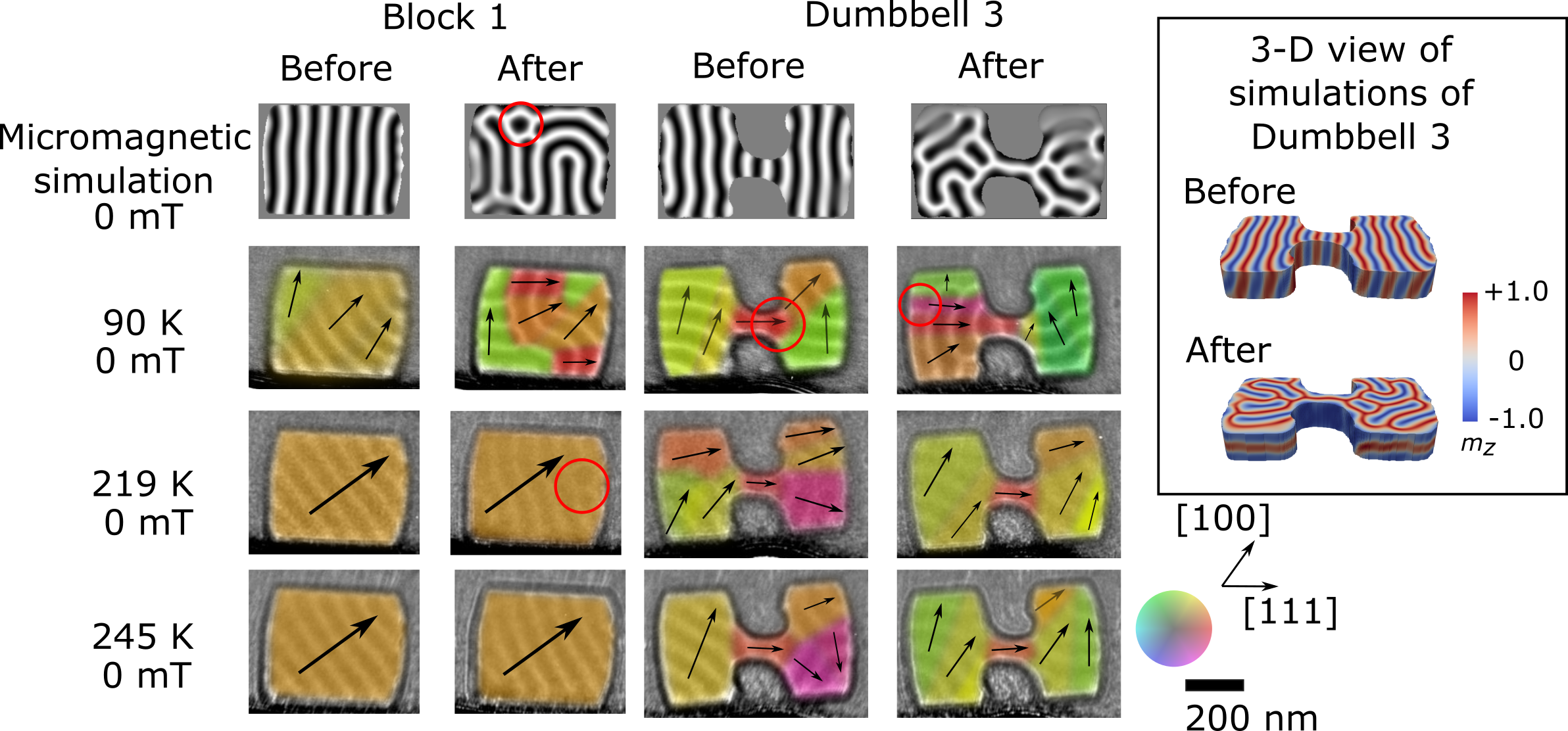}  
    \caption{The magnetic structures formed in the block and dumbbell device-like  shapes are compared using micromagnetic simulations and experimental data acquired at $90$~K, $219$~K and $245$~K before and after an external magnetic field was applied. Under zero applied magnetic field, and at $245$~K the helical phase fills the blocks and dumbbells with a single or slowly varying orientation of the wave vector, ${\bf q}$, whereas at lower temperatures the helical phase forms increasingly more complex, higher energy alignments of magnetisation after the application of an external magnetic field. Metastable skyrmions and short sections of confined helix are marked with red circles. The inset shows a 3D visualisation of the simulated data for dumbbell 3 which corresponds to the helical state before and after the external field was applied, viewed at $30$\textsuperscript{o} to the $[\overline{2}11]$ axis.}
    \label{fig:helicalmontage}
\end{figure}

To understand the magnetic configurations formed in a helimagnet, it is important to consider the history of applied magnetic field and its impact on the observed magnetic phases. Figure~\ref{fig:helicalmontage} shows the helical phase formed in B1 and D3 under zero magnetic field before and after the specimen was subjected to an applied external magnetic field of $313$~mT for the experimental data and $800$~mT in corresponding micromagnetic simulations. (Further details on the micromagnetic simulations are in the Materials and Methods section, including links to the full data set on Zenodo\cite{Zenodo}). Within the simple blocks, under zero field cooling conditions, the magnetic helices form the same pattern irrespective of temperature, with a single wave-vector ${\bf q}$ aligned along $[100]$, as expected from previous low temperature measurements \cite{Lebech89}. At $245$~K the wave-vector alignment along $[100]$ orientation does not change after the application of an external magnetic field.  At $219$~K, the block is mostly filled with helices with a wave-vector alignment along $[100]$, and a metastable skyrmion is present (as marked with a red circle in Figure \ref{fig:helicalmontage}), indicating a slightly increased complexity of the helical phase arrangement after an applied field is removed compared to $245$~K. However, at $90$~K the helices are observed to form a much more complex magnetic structure composed of multiple domains each with a constant wave-vector (as shown by the coloured regions overlaid on $90$~K data in Figure~\ref{fig:helicalmontage}) where the direction of the wave-vector changes abruptly from one domain to the next as the external magnetic field is cycled. 

A more complex magnetic structure is observed at all temperatures in the dumbbell shapes than in the simple blocks. It is particularly interesting to note that the helical wave-vector runs parallel to the edges (along $[111]$) in the central constriction in the dumbbell structure at all temperatures, but outside the constriction, the wave-vector rotates. The higher energy configurations observed at $90$~K in both the blocks and dumbbell shapes which show greater short-range variation in the observed helical ${\bf q}$ vector are consistent with the micromagnetic simulations indicating that they get stuck in metastable configurations and cannot realign into the lowest energy state. Short sections of helices can be seen in the dumbbell shapes as marked with red circles. In this interpretation we have viewed them as helices because we are observing at zero magnetic field, but they could be interpreted as metastable elongated skyrmions. The confined dumbbell geometry is experimentally observed to allow more complex helical states to be stabilised at all temperatures. With a single crystal orientation used for the specimen in this study we cannot deduce whether the alignments of the helical phase wave-vector are controlled predominantly by crystal anisotropy or by geometrical confinement, but through a comparison of the block shapes with the dumbbell shapes we can deduce that a complex device geometry does impact the zero-field micromagnetic alignments present locally in the structure, particularly at temperatures far below the Curie temperature. 

To understand the underlying magnetic structure giving rise to the reduced helical contrast observed in the simulations in dumbbell 3, the 3D simulated magnetisation was examined, as shown in the inset in Figure \ref{fig:helicalmontage}. Before the field was applied, the system was initialised and relaxed with the wave-vector of the helical phase lying in the plane of the sample.  After the magnetic field was applied parallel to the surface normal, the wave-vector of the helical phase can be seen to rotate in some parts of the dumbbell shape to lie out of the sample plane, giving rise to a lower observed contrast in the projected simulations which are used for comparison with the experimental data.  This reduced contrast is also observed in some of the shapes in the experimental data (see full data sets in the Supporting Information).  LTEM is sensitive to the in-plane component of the magnetic flux density averaged through the specimen thickness making it difficult to detect variations in the magnetic structure along the beam direction. Such variations are observed as reductions in contrast in the LTEM image (as seen with skyrmion bobbers \cite{Zheng18}). The full set of experimental data of all of the blocks and dumbbells is included in the Supporting Information and reveals a strong temperature and field dependence on the magnetic structure formed within the shapes.

\subsection{Skyrmion-edge interactions}

\begin{figure}[h!]
    \centering
    \includegraphics[width = 0.9\textwidth]{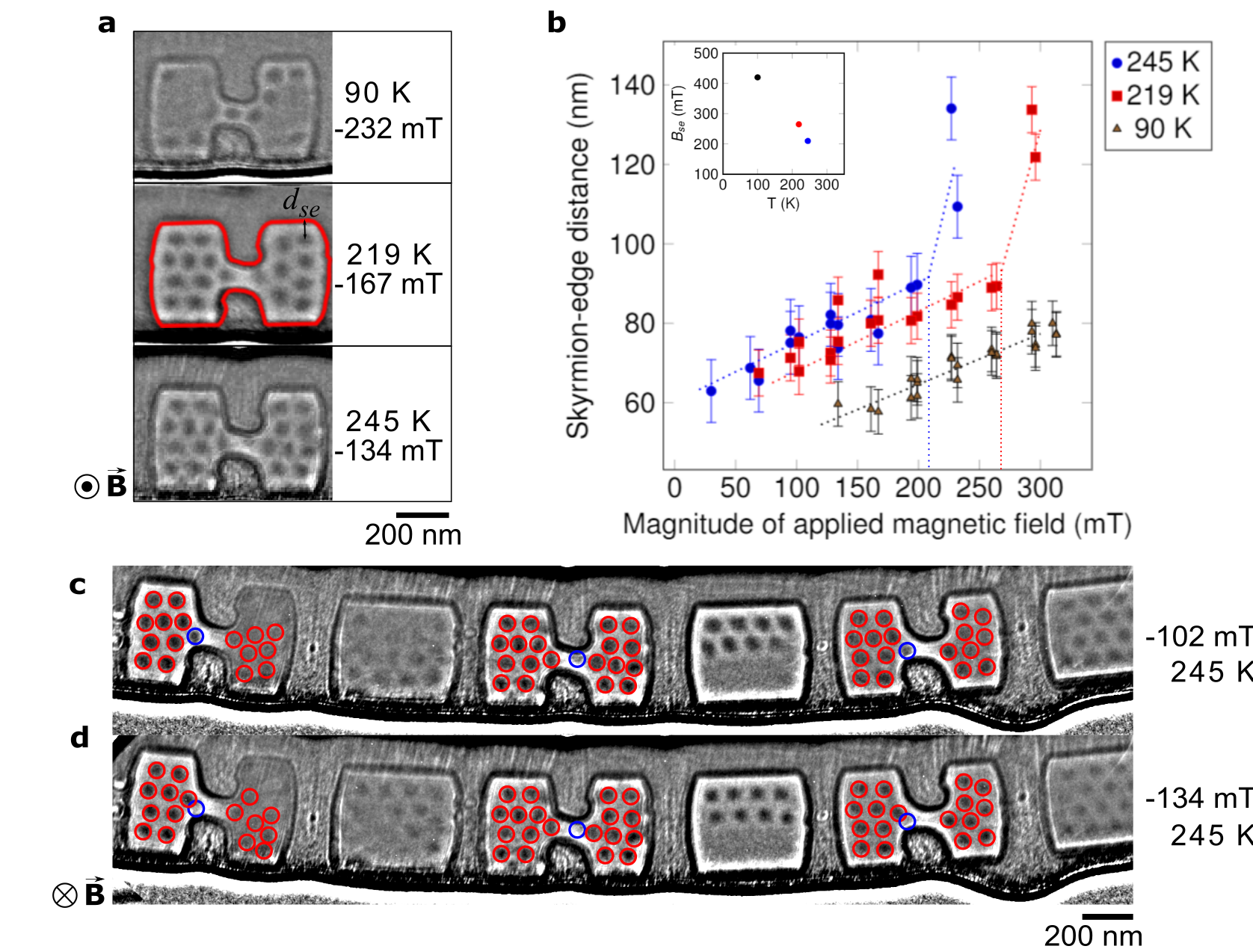}
    \caption{Experimentally observed interaction of skyrmions with edges (a) LTEM images of D2 at the lowest applied magnetic field (when increasing field from 0~mT) with skyrmions in the structure for $90$~K (-$232$~mT), $219$~K (-$167$~mT) and $245$~K (-$134$~mT) (b) Variation of measured skyrmion-edge distance ($d_{\rm se}$) as a function of applied external magnetic field and temperature.  Lines are plotted to reveal the trends of increasing skyrmion-edge distance with magnetic field and temperature.  Error bars represent the mean standard deviation for the skyrmion-edge distances measured at each temperature. Inset graph shows the variation of the critical applied magnetic field, $B_{\rm se}$, at which the skyrmion-edge interaction changes from attractive to repulsive with data from Du {\it et al.}~\cite{Du18} for $100$~K. (c) and (d) Sequential images from the hysteresis loop acquired of the FeGe device-like  shapes at 245 K. With an increase in the (negative) applied external field between (c) $-102$~mT and (d) $-134$~mT the skyrmions in blue circles at the sharp internal corners of the dumbbell shapes disappear. The $70$~nm diameter red circles mark the position of skyrmions that remain in the shapes and move away from the edges as the applied field is increased. The right hand side of dumbbell D1 and block B1 have been significantly affected by strong diffraction contrast which reduces the observed magnetic contrast.}
    \label{fig:edgeposition}
\end{figure}

When confining skyrmions in geometrical structures, edge-skyrmion interactions are increased by the enhanced edge-volume ratio.  Within a confined geometry, it is important to understand any impact there is on skyrmion behaviour under different field and temperature conditions. In order to determine the equilibrium geometrical confinement parameters for these samples, the skyrmion-edge distance, $d_{\rm se}$, was measured for each device-like structure (method described in Materials and Methods). Skyrmions were measured to sit at an equilibrium distance of $d_{\rm se} = 73 \pm 5$~nm from the edge in the block shapes and the dumbbell ends for low applied magnetic fields at $245$~K (as shown in Figure \ref{fig:edgeposition}(a)). Within the central constriction in the dumbbells there is insufficient space for them to sit at this distance from the edge. However, as seen in Figure \ref{fig:edgeposition}(a), skyrmions do form and sit within the constriction indicating that they can be squeezed into confined spaces. This is again temperature-dependent and it is only at $90$~K that a skyrmion is seen at the centre of the constriction, whereas at $219$~K and $245$~K they sit just to the sides of the narrowest constriction. At $90$~K, $d_{\rm se}= 60$~nm but in the constriction, the skyrmion-edge distance is reduced to $47$~nm. Increasing either the temperature or the field causes the skyrmions to move away from the FeGe edges, as shown in Figure \ref{fig:edgeposition}(b) which plots the variation of edge-skyrmion distance, $d_{\rm se}$, as a function of applied field and temperature. This is consistent with predictions by Leonov {\it et al.}~\cite{LeonovPRL16} and experimental data from larger samples by Du {\it et al.}~\cite{Du18} where a linear relationship between $d_{\rm se}$ and applied magnetic field was observed, with a trend for increasing $d_{\rm se}$ as the temperature increased.  The change in gradient of $d_{\rm se}$ as a function of applied magnetic field is theoretically predicted to be at a critical field when the skyrmion-edge interaction changes from being attractive to repulsive \cite{Du18} and the experimental data here shows that the critical field, $B_{\rm se}$, is temperature dependent and reduces as temperature increases (see inset in Figure \ref{fig:edgeposition}b).  

\subsection{Mechanism of skyrmion annihilation}

At $90$~K, skyrmions are still present in the narrow constriction until the applied field exceeds a magnitude of $260$~mT, whereas at $219$~K and $245$~K the equivalent fields are much lower at approximately $160$~mT and $130$~mT. The presence of skyrmions in the narrow region is expected to be dependent on the length and shape of the constriction; here it is short and not parallel-sided. The edge potential acts as a barrier to the skyrmions around most of the sample, and as shown in Figure \ref{fig:edgeposition}, skyrmions move away from edges as the applied field or temperature is increased, as previously noted \cite{Song18}.  However, on detailed examination of the hysteresis loop data acquired at $245$~K as the applied field is increased, some individual skyrmions (marked with blue circles) sitting close to the sharp internal corners of the dumbbell shape are observed to behave differently as shown in Figure \ref{fig:edgeposition} (c) and (d). As the applied field is increased, these marked skyrmions sitting closest to the sharp internal corners are believed to be annihilated through the edge rather than being pushed into the centre, as is observed at all other positions and temperatures in the shapes. This mechanism of annihilation and creation of skyrmions has been proposed and modelled using micromagnetic simulations of a notch \cite{Iwasaki13} and a strip \cite{Cortes17} and has recently been observed in a larger notch geometry \cite{Wang22} than demonstrated here. This behaviour is only observed at $245$~K revealing a thermodynamic dependence to this effect which cannot be modelled in our micromagnetic simulations.  This experimental observation confirms the theoretical predictions that sharp notches or corners can create thermodynamic short-cuts for the annihilation of skyrmions, an effect that has not been previously observed in such small notches.  

\subsection{Optimising skyrmion density}

\begin{figure}[h!]
\centering
\includegraphics[width=16cm]{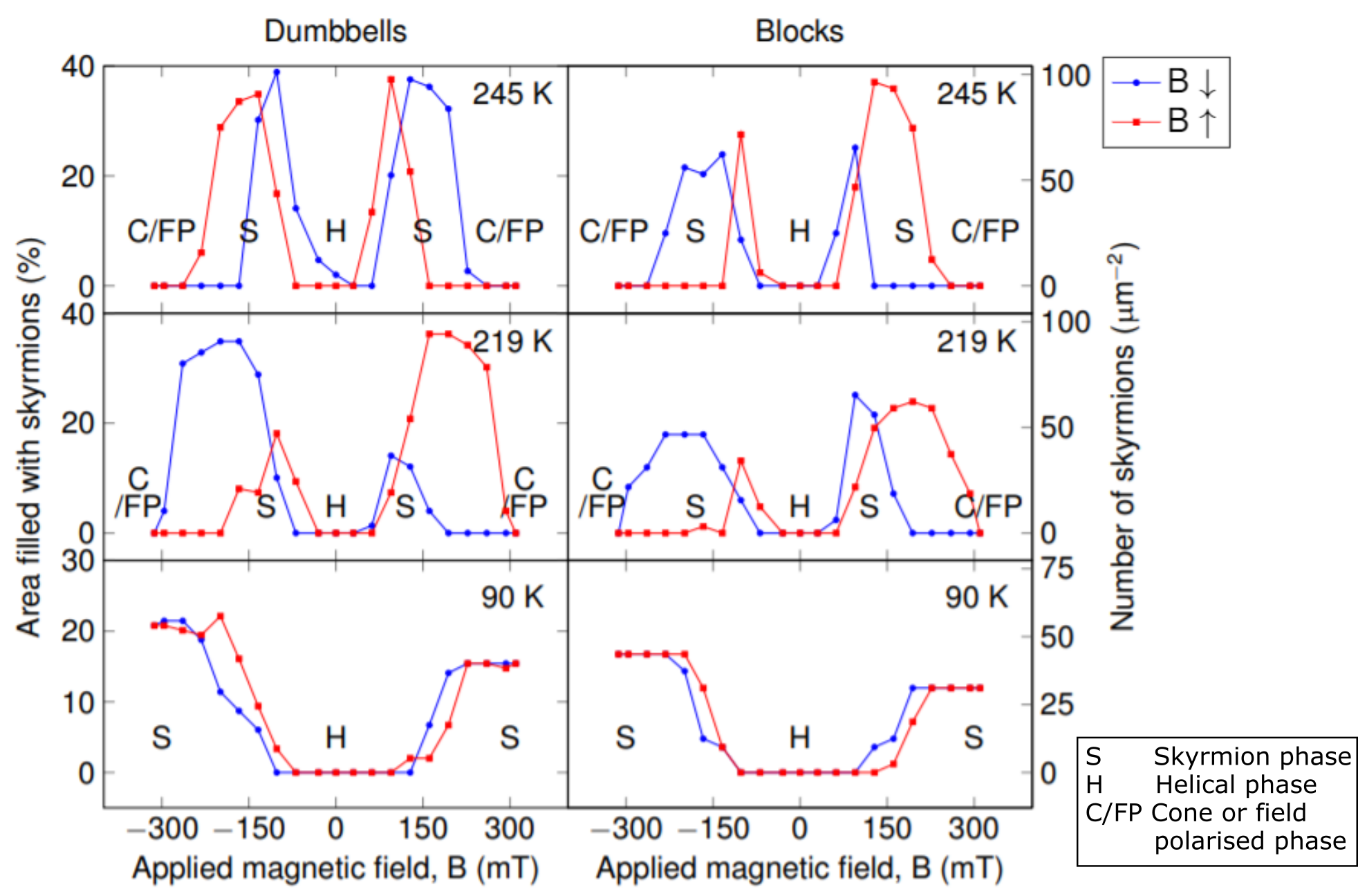}
\caption{Plots showing the percentage area filled with skyrmions (and the number of skyrmions per square micron) in the device-like shapes in all of the blocks and dumbbells as a function of temperature and magnetic field, revealing a strong dependence on direction and magnitude of applied magnetic field (\textbf{B}). At $219$~K we see the most hysteretic behaviour of the device-like shapes revealing that the energy barrier to form skyrmions from the cone phase is much larger than to form skyrmions from the helical phase. These plots reveal that skyrmion formation is enhanced in the dumbbell shapes where the number of skyrmions is higher at all fields and temperatures compared to the block shapes per unit area.}
\label{fig:Bfieldplots}
\end{figure}

Examining the dumbbell and block shapes as a whole, the number of skyrmions formed in all of the shapes (B1--2 and D1--3) as a function of field is shown in Figure \ref{fig:Bfieldplots}. This reveals a strong dependence on temperature and field direction which could be used to optimise skyrmion formation in the device-like shapes. As before, the field loop started at $0$~mT, reduced to $-310$~mT before increasing to $313$~mT and then returning to $0$~mT. At $245$~K, skyrmions are quickly formed as the field strength increases, and retained to higher fields. It is only at $245$~K that thermodynamic barriers are overcome and the hysteretic behaviour of skyrmion formation is reduced whereby equal numbers of skyrmions are present in the device-like shapes when formed from either the field-polarised (or cone) phase or the helical phase. At $219$~K the number of skyrmions formed in the device-like shapes when forming from the helical phase is approximately twice the number that formed from the field-polarised (FP) phase (as seen in the dumbbells plot for $219$~K in Figure \ref{fig:Bfieldplots}), indicating a significant difference in nucleation for skyrmions depending on starting phase (FP or helical) that is not overcome by the increased presence of edges in the device-like shapes. At $90$~K the field-polarised phase was not reached under the conditions used in these experiments and therefore the skyrmions were only observed forming from the helical phase. If we consider the ideal packing of skyrmions within the device-like shapes as a close-packed arrangement, the complex structure of the dumbbell shapes would lead to a lower packing density than the simple block shapes as more space is `wasted'. However, the preferential formation of skyrmions at edges and the space-filling arrangements observed with the skyrmions enhances the skyrmion formation in the complex dumbbell shapes, and a higher skyrmion density is observed in the dumbbells for all temperatures and fields when compared to the simple block shapes which have a similar volume. This quantifies the effect observed earlier, and confirms that increasing the edge surface area to volume ratio can be used to control skyrmion density.

\subsection{Influence of surface damage layer}

These experimental results presented have confirmed that skyrmions have a strong interaction with edges, stabilising the skyrmions at lower fields and allowing a greater density of skyrmions to be formed in the more complex dumbbell shapes than in the simple blocks.  These findings are predicted by simulations \cite{Muller16}, but the exact nature of the specimen surfaces created by FIB milling in these materials is not well defined.  FIB milling is known to form damaged surface layers \cite{McCaffrey01} which are present at all of the edges of these FIB-prepared device-like structures. The thickness of the FIB-modified surface layers was estimated by measuring the extent of the visible darker surface layer present at the edges (evident in Figure \ref{fig:Fig1}d). The measured thickness of these dark surface layers was found to vary around the edges of the device-like structures from 16 to 37~nm (further information on these measurements can be found in the supporting material). Magnetic contrast is observed to continue into these surface layers, visible with the helical phase in Figure \ref{fig:Fig1}(e), and with surface twists in Figure \ref{fig:Fig1}(f)), indicating that some magnetic properties and the crystalline nature of the specimen are preserved close to the specimen surfaces.  Previous studies of FIB-prepared {FeGe} specimens have also shown magnetic contrast very close to the edges of the specimen \cite{Song18, Du18, Jin17, Wolf21}, which is consistent with the experimental results presented in this study.  However, further analysis of these surface layers using more sensitive techniques would reveal the nature of these layers in more detail (for example, high resolution TEM (HRTEM) and energy dispersive x-ray (EDX) mapping).

\section{Conclusions}

In summary, the detailed study of magnetic phases in complex device-like structures has revealed a different behaviour of skyrmions compared to that in a simple geometry. Despite confinement to a length-scale which only allows a single skyrmion within the constriction, skyrmions were still observed to form in the smallest of spaces. These experimental results indicate the significant impact of edges in enhancing skyrmion formation, but the data shows that this is highly temperature and field dependent. The history of specimen temperature and applied external field was revealed to be important when considering the formation of skyrmion and helical phases. The interdependence of skyrmion and helical phase is critical for an understanding of skyrmion formation within a device, and could be exploited to control skyrmion number and position. The strong similarity observed between simulated and experimental results at low temperature indicates that simulations could be used effectively to predict any low-temperature skyrmion behaviour in proposed device geometries but experimental verification is still required for temperatures closer to the Curie temperature.

\section{Materials and Methods}

\subsection{Specimen preparation}
The FeGe single crystals were grown by the chemical vapour transport method with iodine as the transport agent \cite{Richardson67}.  Magnetometry measurements were performed on the bulk FeGe sample. The Curie temperature of the sample, $T$\textsubscript{C}, defined as the point of greatest slope in the magnetisation, $M$, was found to be $280.5$~K. The AC susceptibility was measured during field sweeps after zero field-cooling at each temperature and used to plot the phase diagram shown in Figure \ref{fig:Fig1}(a). A region with a characteristic dip in the AC signal is a well known indicator of the skyrmion lattice state\cite{Bauer12}. Isolated individual dumbbell-shaped and block-shaped lamellae were prepared from a single crystal of FeGe using a FEI Helios Nanolab dual-beam. Initially the $30$~kV Ga\textsuperscript{+} focused ion beam (FIB) was used to prepare a $400$~nm thick lamella still attached at the side and bottom within a trench, and then careful patterning of the device-like shapes using the FIB was carried out to define the narrow constrictions and separations between the dumbbells and block shapes as shown in Figure \ref{fig:Fig1}(c) and in Supporting Information Figure S1. Electron beam platinum deposition was used to coat both sides of the patterned device-like shapes before lifting out the lamella and mounting on a horizontal Omniprobe grid. The grid was rotated to the vertical position and a cross-section of the device-like shapes was prepared $1.8$~\textmu m (as marked by a dotted line in Figure \ref{fig:Fig1}(c)) from the top edge to achieve the correct dimensions of the dumbbell constrictions.

\subsection{Electron Microscopy}

The thinned cross-section of FeGe device-like  shapes was mounted in a Gatan liquid nitrogen cooled model $636$ transmission electron microscope (TEM) holder and examined in an FEI $\text{Titan}^3$ TEM equipped with a Lorentz lens. The specimen was initially mounted in a magnetic field-free condition and a calibrated external magnetic field was applied out of the plane of the specimen using the objective lens of the TEM. To observe magnetic contrast in the TEM, a phase imaging technique is required, and these are only sensitive to the in-plane components of the magnetic field arising from local magnetisation within the specimen. Bloch skyrmions appear as bright or dark areas of contrast when imaged away from focus, depending on defocus and orientation of applied magnetic field. The helical phase can be characterised using defocused imaging, but the field-polarised and cone phases cannot be distinguished from each other as there is no net in-plane component of magnetic field for either when an out-of-plane magnetic field is applied.  LTEM image series were acquired at $90$~K and $263$~K with an applied external magnetic field varying between $0$~mT and $\pm$ $310$~mT. Hysteresis experiments were conducted at $90$~K, $219$~K and $245$~K. The specimen was heated above the Curie temperature after the $90$~K series and before the $219$~K and $245$~K series were acquired, but not between acquisition of the $219$~K and $245$~K series.  LTEM images were acquired on a $2048 \times 2048$ pixel CCD and were energy filtered using a $10$~eV Gatan Tridiem imaging filter. At each change of magnetic field within the hysteresis loop, the acquired LTEM image was refocused before adjusting the defocus to $200$~\textmu m (underfocus) for each image acquired.

\subsection{Micromagnetic Modelling}

The simulations of FeGe geometries of the systems were performed using the software mumax$^3$ \cite{Vansteenkiste2014} in order to try to understand the underlying behaviour of these systems, and to investigate how skyrmions form in `large' confined geometries. In order to do so, we start from the micromagnetic energy density functional $\epsilon[m]$ describing a bulk chiral ferromagnet in the absence of any magnetocrystalline anisotropy:
\begin{equation}
\epsilon[m] = \mathrm{A}(\nabla \mathbf{m}) + \mathrm{D}\mathbf{m}\cdot (\nabla \times \mathbf{m}) - M_\text{s} \mathbf{m}\cdot (\mathrm{H}\mathbf{\hat{e}}_z + \mathbf{H}_\mathrm{d}) 
\end{equation}

\noindent where $\mathrm{H}$ is an externally applied field in the out-of-plane direction, \textbf{m} is the normalised magnetisation and $\mathbf{H}_\mathrm{d}$ is the demagnetising field. In these simulations, we use values for FeGe of saturation magnetisation $M_\text{s} = 384\text{\,kA m}^{-1}$, exchange stiffness $\mathrm{A} = 8.78\,\text{pJ m}^{-1}$, and Dzyaloshinskii-Moriya interaction constant $\mathrm{D} = 1.58\,\text{mJ m}^{-2}$ \cite{Beg15}.

The mask generation process for creating the mesh used in the micromagnetic simulations is described in the Supporting Information Section 4. In the simulations, we initialise each of the five systems with a helical initial state with an in-plane periodicity of $L = 4\pi \mathrm{A} / \mathrm{D}$ and the wave-vector ${\bf q}$ aligned along the $x$-axis. We choose this initial state because it is close to the observed equilibrium experimental systems at $0\,\text{mT}$ far from the Curie temperature. The system is evolved using the steepest descent method \cite{Exl2014} until a metastable equilibrium is reached, which we consider to be when the change in magnetisation, $\lvert\mathrm{d}\mathbf{m}\rvert \leq 10^{-5}$. We then increase the magnetic field from $0$ to $800$\,mT and then reduce it back to $0$\,mT, changing field in steps of $2$\,mT.

\subsection{Image processing}

Diffraction contrast occurs in the LTEM when areas of a specimen are tilted close to a strongly diffracting condition, thereby reducing the intensity observed in that area in a bright field image. In a single crystal specimen it can be reduced by tilting the sample a fraction of a degree, but in this device-like  sample, despite being composed of lamellae cut from an oriented single crystal, there is a small bend which makes it difficult to remove the slowly varying strong diffraction contrast across all of the shapes. No tilt adjustments were made during the acquisition of the hysteresis loops, and therefore to reduce the impact of the diffraction contrast and reveal positions of the magnetic features more clearly, the aligned experimental images were enhanced using a high-pass filter and subsequent equalising the standard deviation across a series of images. The full unprocessed and processed data sets are shown in the Supporting Information Figures S3-8. Areas that have been strongly affected by diffraction contrast show reduced magnetic contrast, for example, in the right hand side of D1 from $-69$\,mT to $-199$\,mT. When the applied field direction is reversed in the hysteresis loop, the skyrmion contrast also reverses - i.e. skyrmion cores initially appear dark and change to appearing bright. The chirality of the skyrmions is preserved but their core magnetisation is reversed and therefore the contrast observed in the TEM is reversed. The transport of intensity equation (TIE) \cite{Beleggia04} was used to reconstruct the in-plane component of the magnetic flux density from the defocused images. For a one-sided TIE reconstruction, a single defocused image was used to reconstruct the flux density as described by Chess \emph{et al.} \cite{Chess17}. This technique is very sensitive to all changes in both magnetic and electrostatic potential and therefore significant edge effects are also observed at the sharp change in mean inner potential at the {FeGe}-{Pt} interfaces. For a more quantitative analysis of the phase variation arising from skyrmions in these device-like  shapes, off-axis electron holography is required where a lower spatial resolution is obtained but any edge effects are not delocalised by the impact of defocus.

The skyrmion-edge distance was measured by creating masks of the FeGe shapes, as for the micromagnetic simulations (see Supporting Information), and applying the masks to each of the defocused images in each hysteresis loop. The masks were rescaled using the calibration of the magnification change from in-focus to the defocus of the hysteresis loop. An accurate measurement of skyrmion centre to edge distance was obtained from a normal drawn between the defined shape edge and the centre of the skyrmion.

\section{Acknowledgements}

This work was supported by the UK Skyrmion Project EPSRC Programme Grant (EP/N032128/1). R.A.P acknowledges funding from EPSRC's Centre for Doctoral Training in Next Generation Computational Modelling, (EP/L015382/1). R.A.P and H.F. acknowledge the use of the IRIDIS High Performance Computing Facility, and associated support services at the University of Southampton, in the completion of this work. Contributions of J.A.T. Verezhak and A. \v{S}tefan\v{c}i\v{c} to the crystal growth work at the University of Warwick are gratefully acknowledged.

The micromagnetic simulation data that support the findings of this study are available in Zenodo https://doi.org/10.5281/zenodo.4270366. The experimental data supporting the findings of this study are available within the paper (and its Supporting Information files).

Data in Figure 1(a) and 1(b) was reprinted (Adapted or Reprinted in part) with permission from 'Real-space imaging of confined magnetic skyrmion tubes', M. T. Birch \emph{et al.}, Nature Communications, 11, 1726. Copyright 2020 Springer Nature.

\section{Supporting Information}
Details of sample synthesis and preparation, electron microscopy, FIB damage of specimen surfaces, mask generation for micromagnetic simulations, Secondary electron images of sample preparation process with schematic of finished specimen, LTEM image and line scan with table of surface damage layer measurements, table of perimeter and area of shapes, LTEM montage images of unprocessed and processed data sets from hysteresis loops at $90$, $219$ and $245$~K. 

\bibliography{main}
\newpage
\subsection{For Table of Contents Use Only}
\begin{figure}
    \centering
    \fbox{\includegraphics{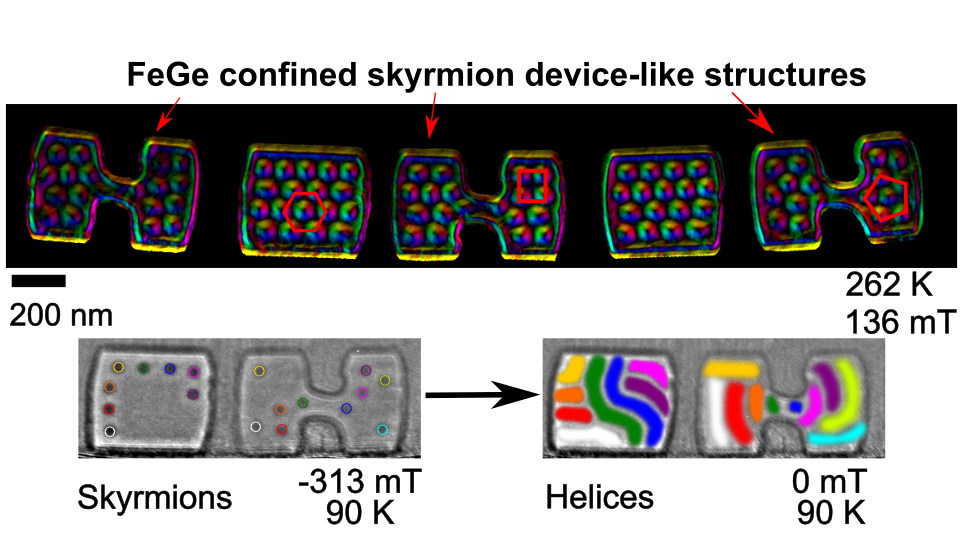}}
    \label{fig:TOC}
\end{figure}
\end{document}